# Pulsed-mode metalorganic vapor-phase epitaxy of GaN on graphene-coated *c*-sapphire for freestanding GaN thin films

*Seokje Lee[1], Muhammad S. Abbas[2], Dongha Yoo[1], Keundong Lee[3], Tobiloba G. Fabunmi[1], Eunsu Lee[1], Imhwan Kim[1], Daniel Jang[4], Sangmin Lee[5], Jusang Lee[5], Ki-Tae Park[6], Changgu Lee[4,7], Miyoung Kim[5], Yun Seog Lee[6], Celesta S. Chang[1], Gyu-Chul Yi[1]\**

AUTHOR ADDRESS

[1]Department of Physics and Astronomy, Seoul National University, Seoul 08826, Republic of Korea

[2]Department of Physics, Sungkyunkwan University College of Natural Science, Suwon 16419, Republic of Korea

[3]Department of Electrical and Computer Engineering, University of California San Diego, La Jolla, CA 92093, United States

[4]SKKU Advanced Institute of Nano Technology, Sungkyunkwan University, Suwon 16419, Republic of Korea

[5]Department of Materials Science and Engineering, Seoul National University, Seoul 08826, Republic of Korea

[6]Department of Mechanical Engineering, Seoul National University, Seoul 08826, Republic of Korea

[7]School of Mechanical Engineering, Sungkyunkwan University College of Engineering, Suwon 16419, Republic of Korea

*Corresponding author: gcyi@snu.ac.kr

ABSTRACT

We report the growth of high-quality GaN epitaxial thin films on graphene-coated *c*-sapphire substrates using pulsed-mode metalorganic vapor-phase epitaxy, together with the fabrication of freestanding GaN films by simple mechanical exfoliation for transferable light-emitting diodes (LEDs). High-quality GaN films grown on the graphene-coated sapphire substrates were easily lifted off using thermal release tape and transferred onto foreign substrates. Furthermore, we revealed that the pulsed operation of ammonia flow during GaN growth was a critical factor for the fabrication of high-quality freestanding GaN films. These films, exhibiting excellent single crystallinity, were utilized to fabricate transferable GaN LEDs by heteroepitaxially growing $In_xGa_{1-x}N$/GaN multiple quantum wells and a *p*-GaN layer on the GaN films, showing their potential application in advanced optoelectronic devices.

Manufacturing high-quality inorganic thin films on large-area flexible substrates is a key technology required for flexible devices such as wearable sensors,[1] flexible displays,[2,3] and electronic devices.[4] One promising approach for the fabrication of flexible devices using inorganic thin films is to grow epitaxial thin films on two-dimensional (2-D) layered materials such as graphene and *h*-BN since the 2-D materials offer weak bonding with as-grown films for easy exfoliation of epilayers from 2-D materials.[2,5-8] In particular, an epitaxial method called remote epitaxy enables us to grow single-crystalline thin films on graphene-coated polar substrates where the partially screened electrostatic potential of the substrate allows the epitaxial relationship between the substrate and the epilayer.[9,10] Recent studies have demonstrated successful remote epitaxy of single-crystalline thin films for various III-V,[9,11] III-N,[12-14] and complex-oxide materials, showing its potential for providing flexible and high-quality materials platforms for future device applications.[15-18]

While the remote epitaxy of single-crystalline GaN thin films has also been studied, its growth has mainly been limited to molecular beam epitaxy (MBE).[12,19] However, considering the ultra-high vacuum required for MBE, limited sizes of GaN films, and low throughput, more improvements are expected to reach the commercialization of remote epitaxially grown GaN films. Meanwhile, metalorganic vapor-phase epitaxy (MOVPE) is an epitaxial method commercially used in industrial settings with its ability for wafer-scale manufacturing, promising for mass production of high-quality GaN devices. Although remote epitaxy of other III–V materials such as GaAs, InP, and GaP using MOVPE has already been reported,[9,11] no reports were made for GaN thin films that show complete exfoliation from graphene-coated substrates, which is one of the prerequisites to acknowledge as successful remote epitaxy. This is due to $H_2$ and $NH_3$ ambient gases that degrade graphene at high growth temperatures of GaN MOVPE, which inevitably leads to failure in GaN exfoliation from the graphene-coated substrate.[20-22]

To circumvent this problem, a more robust graphene-like buffer layer (GBL) obtained through the graphitization of SiC instead of graphene has been used for the remote epitaxy of GaN thin films in MOVPE.[23] Alternatively, remote epitaxial growth and exfoliation of GaN micro-rods, which require a relatively low gas flow rate and growth temperature compared to those of GaN thin films, have also been reported.[24] However, these approaches do not directly solve the inherent problem, while the range of substrate and application field still remains limited. To enable mass production of high-quality GaN-based devices, developing a simple and universal recipe for remote epitaxy of GaN using MOVPE would be crucial.

Here, we report a strategy for MOVPE growth of transferable single-crystalline GaN thin films on graphene-coated sapphire substrates by using the pulsed operation of $NH_3$ flow to prevent the deterioration of graphene substrates during GaN growth. The $NH_3$ pulsed-flow MOVPE technique, where the $NH_3$ injection to the reactor is periodically interrupted, was used to enhance the lateral growth of the GaN buffer layer by a controlled V/III ratio.[25-27] The continuous buffer layer created by lateral overgrowth was shown to protect graphene from damage under high-temperature GaN growth, which we confirmed by various characterization methods and successful exfoliation of high-quality GaN thin films without even using a metallic stressor layer. The quality of the GaN film was demonstrated by fabricating light-emitting diodes (LEDs) and transferring them onto a copper foil, showing the potential application of remote-epitaxial GaN films grown by MOVPE.

RESULTS AND DISCUSSION

The basic strategy for epitaxial growth of transferable single-crystalline GaN thin films on graphene-coated *c*-sapphire substrates is shown in Figure 1. As a template for GaN growth, graphene was directly grown on a *c*-sapphire substrate by chemical vapor deposition to eliminate the graphene transfer process that inevitably introduces unwanted defects and residues.[28,29] The graphene-coated sapphire substrates were treated using oxygen plasma in order to enhance the nucleation of GaN on the surface of graphene (Figure 1a),[2,30] which we confirmed with Raman spectroscopy and an exfoliation test (Figure S1). To obtain high-quality GaN thin films with a uniform and smooth surface, we developed a three-step growth method that increases the nucleation sites while preventing the deterioration of graphene. In the first growth step, a 150-nm thick low-temperature GaN (LT-GaN) buffer layer was grown at 600 °C for 10 min (Figure 1b). For the second growth step, we used a pulsed epitaxial lateral overgrowth (PLOG) method at a higher temperature of 1050 °C for 10 min, where the $NH_3$ flow was periodically interrupted while trimethylgallium (TMGa) was continuously injected into the reactor (Figure 1c). Under Ga-rich conditions enabled by the pulsed $NH_3$ injection, Ga adatoms tend to laterally diffuse towards $\{10\bar{1}1\}$ and $\{10\bar{1}0\}$ facets.[27,31] This promoted the lateral growth of the GaN buffer, resulting in the formation of a continuous film over graphene.[32] Additionally, employing the $NH_3$ pulsed-flow growth could mitigate damage to the graphene by effectively reducing the $NH_3$ partial pressure since we observed that annealing under a high-temperature $NH_3$ environment severely etched the $O_2$ plasma-treated graphene (Figure S2). The continuous PLOG layers protected the graphene from damage at the next stage. Additional layers were grown on top of the PLOG layers at a higher temperature (1150 °C) for 30 min under a continuous flow of $NH_3$ to obtain high-quality GaN thin films (Figure 1d).

A series of GaN growth experiments were conducted with different NH$_3$ flow modes at the second growth step to investigate the influence on the surface morphology and exfoliation of GaN films grown on graphene-coated sapphire substrates. In the second growth step at 1050 °C, various NH$_3$ flow modes were individually tested for the same growth time of 10 min, including continuous-flow mode, 3s-on/1s-off, 2s-on/2s-off, and 1s-on/3s-off injection/interruption pulsed-flow modes. Throughout these experiments, the first growth step at 600 °C and the third growth step at 1150 °C maintained a continuous NH$_3$ flow. When NH$_3$ was continuously supplied during the second growth step, GaN islands were not merged, resulting in discontinuous surface morphology, as shown in the scanning electron microscopy (SEM) image (Figure 2a). In contrast, a 3s-on/1s-off pulsed flow of NH$_3$ during the second growth step led to merged GaN surfaces with 200-400 nm diameter pinholes (Figure 2b). A 2s-on/2s-off pulsed flow of NH$_3$ produced a fully merged and continuous GaN film (Figure 2c). This enhanced lateral growth under the NH$_3$ pulsed-flow mode was attributed to Ga adatom lateral diffusion to $\{10\bar{1}1\}$ and $\{10\bar{1}0\}$ facets in Ga-rich conditions during NH$_3$ interruption time. However, when the NH$_3$ injection time was significantly shorter than the interruption time, as in the 1s-on/3s-off mode, excess Ga adatoms accumulated, resulting in a rough surface morphology (Figure 2d). After the third growth step at 1150 °C with continuous NH$_3$ flow, all samples became smoother and more continuous than those seen after the second growth step (Figure 2e-h). Nonetheless, pinholes remained visible on the film surfaces of samples with incomplete GaN merging after the second growth step, such as those grown using NH$_3$ continuous-flow and 3s-on/1s-off pulsed-flow modes (Figure 2e,f). Meanwhile, the samples grown with 2s-on/2s-off and 1s-on/3s-off NH$_3$ pulsed-flow modes in the second growth step exhibited featureless and flat surfaces after the third growth step (Figure 2g,h).

We also attempted to exfoliate GaN films from the graphene-coated sapphire substrates using thermal release tape. Notably, films grown with $NH_3$ continuous-flow and 3s-on/1s-off pulsed-flow modes were not exfoliated (Figure 2i,j). In contrast, over 90% of the films grown with 2s-on/2s-off and 1s-on/3s-off $NH_3$ pulsed-flow modes were successfully exfoliated using only the thermal release tape (Figure 2k,l). To confirm whether the film simply needs higher stress for exfoliation, a metallic stressor layer was deposited before applying thermal release tape. However, GaN grown on graphene-coated sapphire under $NH_3$ continuous-flow mode in the second growth step were still unable to be exfoliated, while the GaN grown by 2s-on/2s-off pulsed-flow mode showed 99% exfoliation of the film (Figure S3). Additionally, Raman spectroscopy confirmed the presence of graphene on the surface of the original substrate after GaN exfoliation (Figure S4). We speculate that the $NH_3$ pulsed-flow growth not only enhanced the lateral growth of the buffer layer but also mitigated $NH_3$-induced damage to graphene by effectively reducing the $NH_3$ partial pressure in the reactor during the second growth step. As a result, the continuous buffer layer grown in the second growth step with the pulsed-flow $NH_3$ led to smooth and continuous GaN films that could be exfoliated after the final growth step.

The GaN/graphene/sapphire interfaces of two individual GaN samples, one grown with $NH_3$ continuous-flow mode and the other with the 2s-on/2s-off pulsed-flow mode, were investigated using cross-sectional scanning transmission electron microscopy (STEM). As shown in a representative high-angle annular dark field (HAADF) STEM image of the GaN sample grown under the $NH_3$ continuous-flow mode (Figure 3a), discontinuous graphene between GaN and sapphire was observed. GaN appeared to have grown directly on sapphire through a locally damaged graphene area which is likely to be the main reason why the sample could not be exfoliated. Moreover, especially in the region where graphene is damaged, the atomic arrangement of GaN is not precisely aligned with the *c*-axis (0001) but tilted with respect to the *c*-axis (Figure 3b). This tilting of GaN could be caused by lateral growth of GaN along irregular surfaces of damaged graphene, where pits or steps can be created under the continuous flow of $NH_3$ at the second growth step.[33] In contrast, a clear separation of GaN and sapphire by a 6 Å graphene gap is shown in a representative HAADF-STEM image of the GaN sample grown under the 2s-on/2s-off pulsed flow of $NH_3$ at the second growth step (Figure 3c), which means that the layered structure of graphene was well preserved during the lateral growth of GaN under the pulsed flow of $NH_3$. In addition, GaN shows a well-aligned growth direction parallel to the *c*-axis (Figure 3d). The only difference between these two GaN samples was whether the $NH_3$ was supplied in the continuous or pulsed flow mode at the second growth step among the three growth steps. Hence, these distinctly different GaN/graphene/sapphire interfaces of the two samples support that the $NH_3$ pulsed-flow growth at the second growth step has significantly reduced graphene damage leading to improved crystalline quality of GaN.

We also investigated the heteroepitaxial relationship of GaN thin films grown on graphene-coated *c*-sapphire using the three-step growth method with 2s-on/2s-off $NH_3$ pulsed-flow mode during the second growth step. A selected area electron diffraction (SAED) pattern was taken from the GaN/graphene/sapphire interface, as shown in the low-magnification cross-sectional STEM image of Figure 4a. SAED pattern reveals heteroepitaxial relationship of GaN (0002) // sapphire(0006) and GaN(11$\bar{2}$0) // sapphire(30$\bar{3}$0) between GaN and sapphire (Figure 4b). The same epitaxial relationship was confirmed in the SAED pattern of GaN on bare *c*-sapphire grown in the same batch with GaN on graphene-coated *c*-sapphire (Figure S5). This indicates that GaN thin films were grown, maintaining a heteroepitaxial relationship with the *c*-sapphire substrate despite the presence of graphene between GaN and *c*-sapphire. Two possible mechanisms can explain this heteroepitaxial relationship across 2-D materials. One is remote epitaxy, and the other is pinhole-seeded epitaxy where GaN nucleates directly from the bottom substrate through pinholes on 2-D materials and laterally overgrows to form a continuous film.[34] By investigating the entire 4.4-μm long cross-sectional region of the GaN films, we could not find any region where GaN was directly connected to sapphire through the graphene layers. This result suggests that the heteroepitaxial growth does not result from pinhole-seeded epitaxy (Figure S6).

The crystallinity and growth orientation of GaN thin films grown on graphene-coated *c*-sapphire were further investigated by electron backscatter diffraction (EBSD). As shown in Figure 4c, the uniform red color in the normal Z-direction inverse pole figure (IPF) map shows that the GaN thin films were well-aligned along the *c*-axis. Furthermore, IPF maps in the X and Y directions exhibit uniform green and blue colors, respectively, which indicate that the GaN films were epitaxially grown in a single in-plane domain. X-ray diffraction (XRD) measurements also revealed the single crystallinity of GaN films grown on graphene/*c*-sapphire. XRD theta-2theta spectrum shows strong intensity peaks of GaN(0002) and GaN(0004) at 34.60° and 72.93°, respectively, indicating GaN film with a wurtzite structure grown along the (0001) direction (Figure 4d). The GaN (0002) rocking curve, with a full-width half maximum (FWHM) value of 0.19°, indicates that GaN thin films were well aligned along the *c*-axis (Figure 4e). In addition, the azimuthal scan demonstrates in-plane six-fold rotational symmetry for the GaN films (Figure 4f).

To demonstrate the optoelectronic application of our transferable single-crystalline GaN thin films, we fabricated GaN thin-film LEDs before exfoliation and then transferred them onto Cu foils. The schematic diagram illustrating the LED fabrication process is shown in Figure 5a. Initially, Si-doped *n*-GaN, ten-period $In_xGa_{1-x}N$/GaN MQWs, and Mg-doped *p*-GaN layers were deposited on the PLOG-GaN grown on a graphene-coated sapphire substrate. Doped and multi-quantum-well (MQW) layers were grown under an NH$_3$ continuous-flow condition to minimize defect formation. For the electroluminescent (EL) device fabrication, a 20 nm/20 nm Ni/Au bilayer was deposited on the *p*-GaN surface. Next, thin-film LEDs were exfoliated from the graphene-coated sapphire substrate using a thermal release tape. A 20 nm/20 nm In/Au bilayer was deposited on the exposed *n*-GaN surface after exfoliation. Finally, thin-film LEDs were transferred onto Cu foil using a conductive tape with the *n*-GaN surface facing the foil.

At an applied bias voltage of 6 V, the thin-film LEDs on Cu foil exhibited strong blue light emission that could be seen under normal room illumination (Figure 5b). LED characteristics were investigated by measuring the current-voltage characteristic (*I-V*) curve and EL spectra. The *I-V* characteristic curve of LEDs shows a typical rectifying behavior of GaN p-n junction diodes (Figure 5c). EL spectra were obtained at various applied bias voltages ranging from 3.5 V to 6 V (Figure 5d). At an applied bias voltage of 3.5 V, the dominant EL peak was observed at 486 nm, which shifted to 468 nm as the applied bias voltage increased to 6 V. The change in the EL peak position could be presumably due to the band-filling effect of the localized energy states, commonly observed in LEDs with GaN/In$_x$Ga$_{1-x}$N MQWs.[35]

CONCLUSION

In summary, we introduce a new strategy to enable high-quality GaN remote epitaxy with MOVPE, by simply controlling the NH$_3$ flow at a certain GaN growth stage. Under the pulsed flow of NH$_3$ with a controlled V/III ratio, we observed that the lateral growth of the GaN buffer layer on graphene was greatly enhanced, which later served as a protection layer for graphene preventing its damage under high-temperature GaN growth. Consequently, we obtained high-quality GaN thin films grown by remote epitaxy on graphene-coated *c*-sapphire substrates and demonstrated successful exfoliation of these films using only thermal release tape. The crystallinity and quality of the GaN film were confirmed by cross-sectional TEM, EBSD, and XRD characterization, followed by a demonstration of LEDs fabricated using the GaN film grown by our method. This study presents a straightforward approach to reducing the 2-D material damage caused by gas flow in the harsh GaN MOVPE environment, opening up new possibilities for mass production of high-quality GaN devices and other III–N or III–V material growth on graphene. Therefore, we expect our study will pave the way for various applications and highly efficient GaN device production in bioelectronics and flexible displays.

METHODS

**Direct growth of graphene on sapphire substrates.** Graphene was grown on *c*-sapphire using a catalyst-free chemical vapor deposition (CVD) method. 2-inch *c*-sapphire substrates were put in a CVD furnace, purged with Ar for 10 minutes, and heated to 1160 °C. A mixture of Ar, $H_2$, and $CH_4$ gases was introduced into the furnace to synthesize graphene on *c*-sapphire substrates at 20 torr for 1 hour. The furnace was then rapidly cooled with Ar flow.

**MOVPE growth of GaN thin films on graphene-coated sapphire substrates.** Graphene-coated *c*-sapphire substrates underwent a 3-sec $O_2$ plasma treatment before entering a GaN MOVPE chamber. Using a horizontal MOVPE system, GaN thin films were grown with TMGa and $NH_3$ as precursors. The chamber had a consistent 100 torr pressure. The substrate was cleaned and annealed in an $H_2$ and $NH_3$ mix at 800 °C for 10 min. A 150-nm LT-GaN buffer layer was applied at 600 °C. PLOG merged the LT-GaN layer with pulsed $NH_3$ flow in 150 cycles. During PLOG, a 200 sccm $NH_3$ flow was periodically injected (on) into the reactor for 2 sec and interrupted (off) for 2 sec in one cycle, while TMGa carried by 4 sccm $H_2$ was continuously introduced into the reactor. Finally, HT-GaN was grown at 1150 °C for 30 min, resulting in ~2.5 μm thick GaN films.

**GaN thin-film LED fabrication.** To make transferable GaN LEDs, a 1.5 μm Si-doped n-GaN was regrown on a 1 μm PLOG GaN layer on graphene-coated sapphire substrate. Ten periods of $In_xGa_{1-x}N$/GaN MQWs followed, with 5 nm GaN barriers and 3 nm $In_xGa_{1-x}N$ wells. A 100 nm Mg-doped *p*-GaN layer completed the LED structure. As-grown GaN LEDs underwent thermal activation at 750 °C for 10 minutes. A Ni/Au (20 nm/20 nm) bilayer was deposited to the *p*-GaN and annealed at 500 °C for 5 min. The entire LED structure was released by a thermal release tape from the graphene substrate, and a 20 nm/20 nm In/Au bilayer was deposited on the *n*-GaN. Finally, the LED structure was transferred onto a conducting carbon tape/Cu foil, with the *n*-GaN surface facing the foil.

**Characterization.** The surface morphology of GaN films was studied using a Zeiss field-emission scanning electron microscope (FE-SEM) (Merlin Compact). Their crystal structure was analyzed using a PANalytical X-ray diffractometer with Cu Kα radiation. Crystal orientation was checked with an EBSD detector (Oxford Instruments) on a JSM-7600F FE-SEM. Cross-sections of GaN films were made using a Helios G4 focused ion beam (FIB) system with Ga+ ions. GaN/graphene/sapphire interfaces were imaged at atomic resolution with a Themis Z STEM system (Thermo Fisher Scientific). For LED testing, a Keithley 2400 source meter measured current-voltage curves and applied DC voltages. EL spectra were recorded with an Andor Technology CCD (DU401A) linked to a Dongwoo Optron monochromator (DM150i).

FIGURES

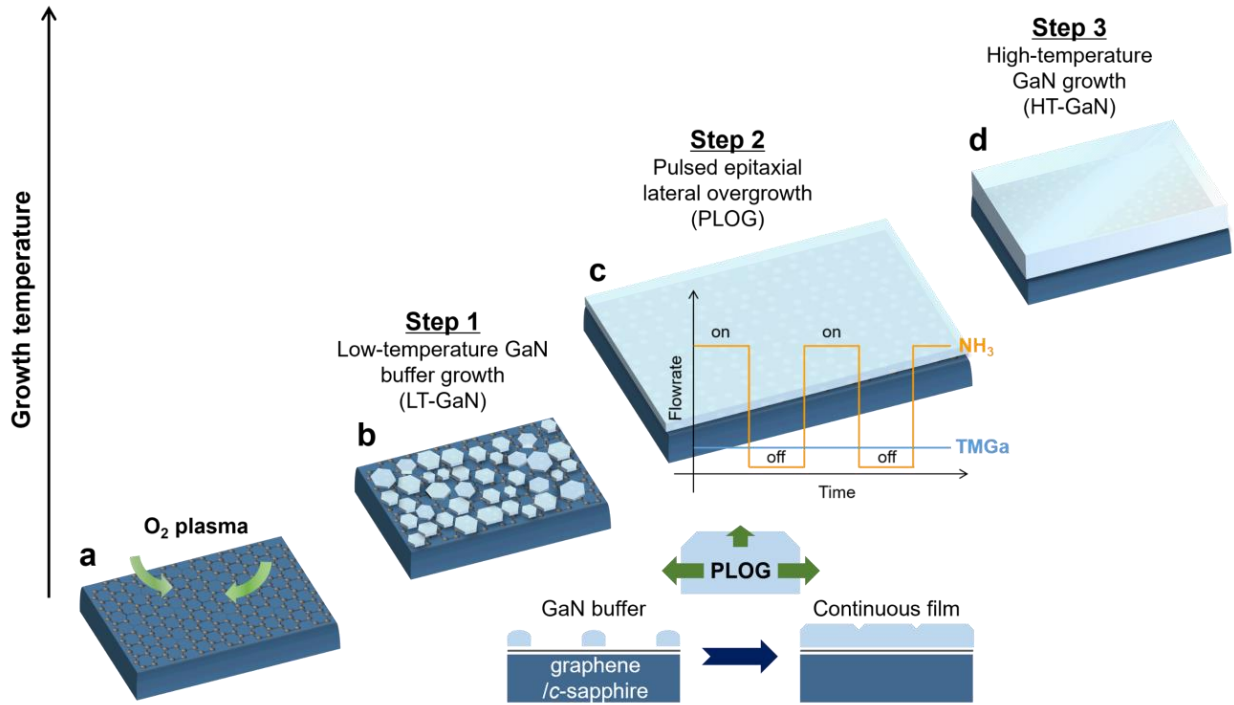

**Figure 1.** Schematic of a three-step growth process for high-quality GaN thin films on a graphene-coated *c*-sapphire substrate using the $NH_3$ pulsed-flow growth technique. The process includes (a) $O_2$ plasma treatment to increase nucleation sites on graphene, (b) low-temperature GaN (LT-GaN) buffer layer growth, (c) pulsed epitaxial lateral overgrowth (PLOG) under a pulsed flow of $NH_3$ to form a continuous buffer layer over the graphene, and (d) high-temperature GaN (HT-GaN) growth under a continuous flow of $NH_3$ to obtain high-quality GaN thin films.

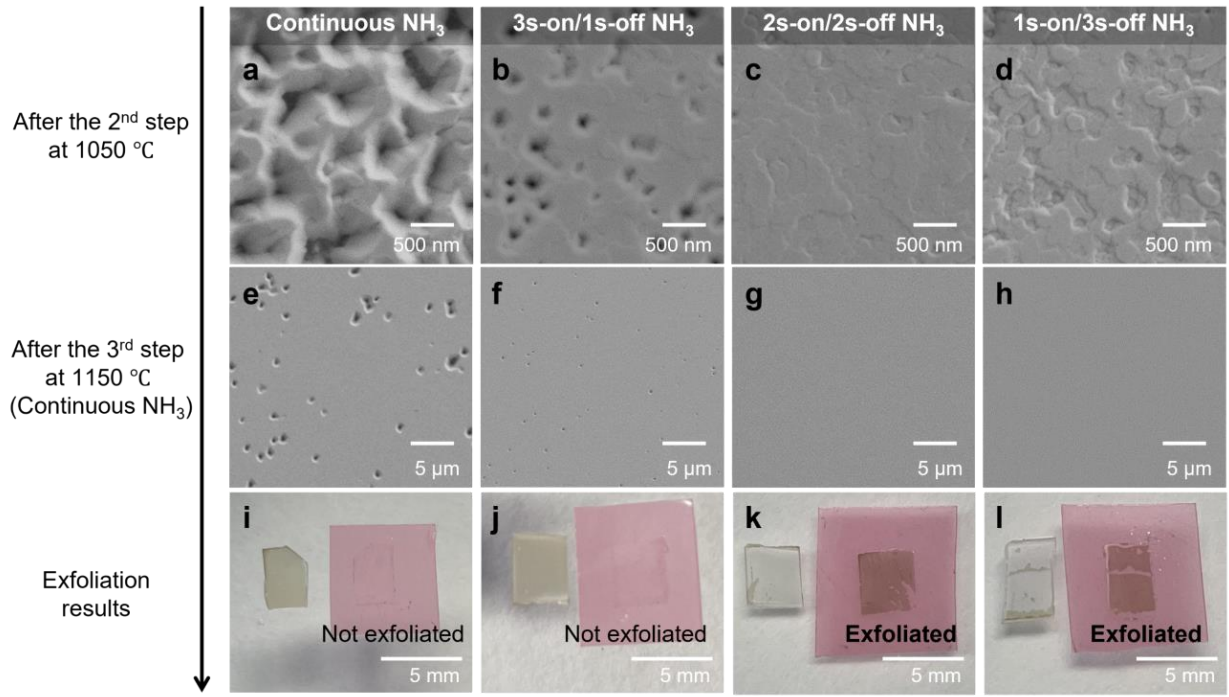

**Figure 2.** SEM images and exfoliation results of GaN films grown on graphene-coated *c*-sapphire substrates using various NH$_3$ flow modes in the second growth step at 1050 °C. (a-d) SEM images of GaN samples after the second growth step with different NH$_3$ flow modes: (a) continuous-flow, (b) 3s-on/1s-off pulsed-flow, (c) 2s-on/2s-off pulsed-flow, and (d) 1s-on/3s-off pulsed-flow mode samples. (e-h) SEM images of GaN samples after the third growth step at 1150 °C: (e) continuous-flow, (f) 3s-on/1s-off pulsed-flow, (g) 2s-on/2s-off pulsed-flow, and (h) 1s-on/3s-off pulsed-flow mode samples. (i-l) Photographs showing the specimen and the exfoliation results for GaN samples using thermal release tape: (i), (j) shows no exfoliation, whereas (k) and (l) shows clear substrate and the exfoliated films on the thermal release tape side by side.

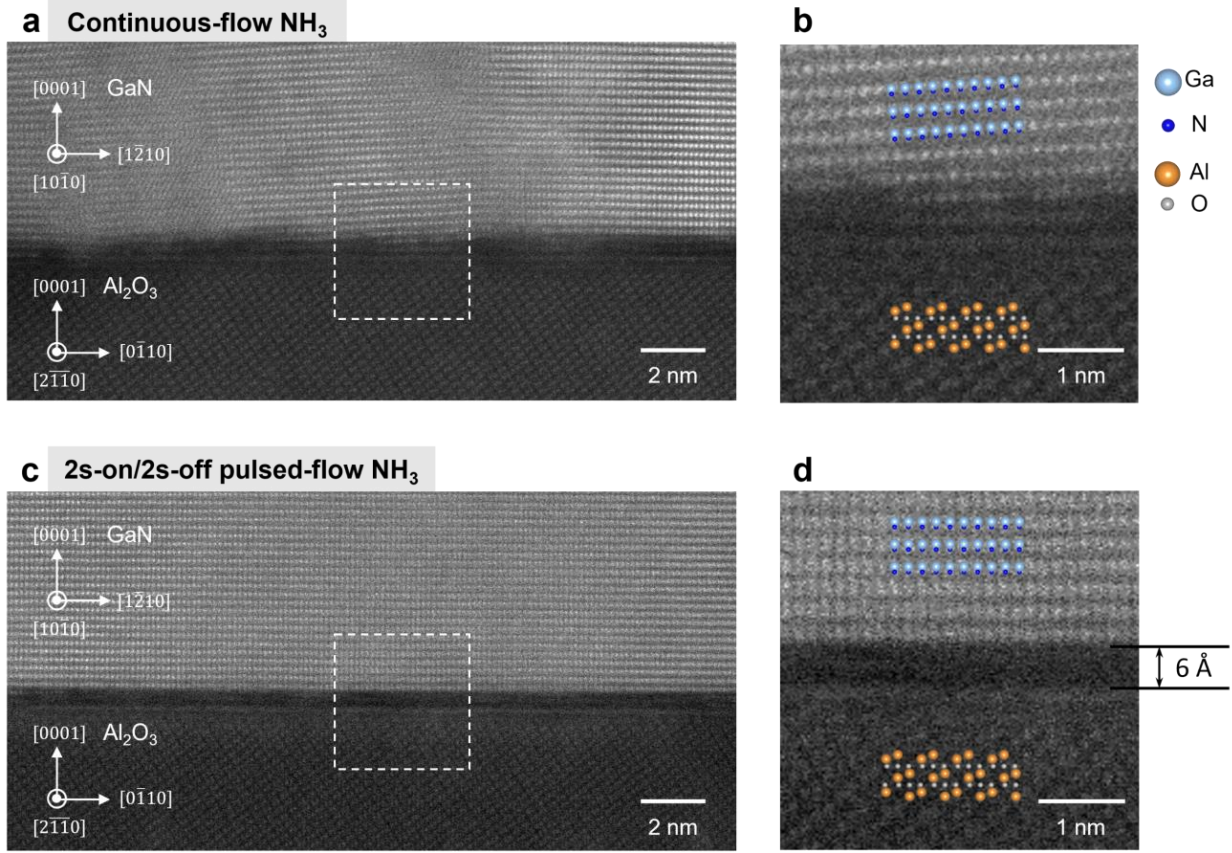

**Figure 3.** (a) Cross-sectional STEM image of GaN thin films on graphene-coated *c*-sapphire substrate grown using continuous-flow $NH_3$ in the second growth step and (b) a magnified image of the dashed box. (c) Cross-sectional STEM image of GaN thin films on graphene-coated *c*-sapphire substrate grown using a 2s-on/2s-off pulsed-flow $NH_3$ in the second growth step and (d) a magnified image of the dashed box.

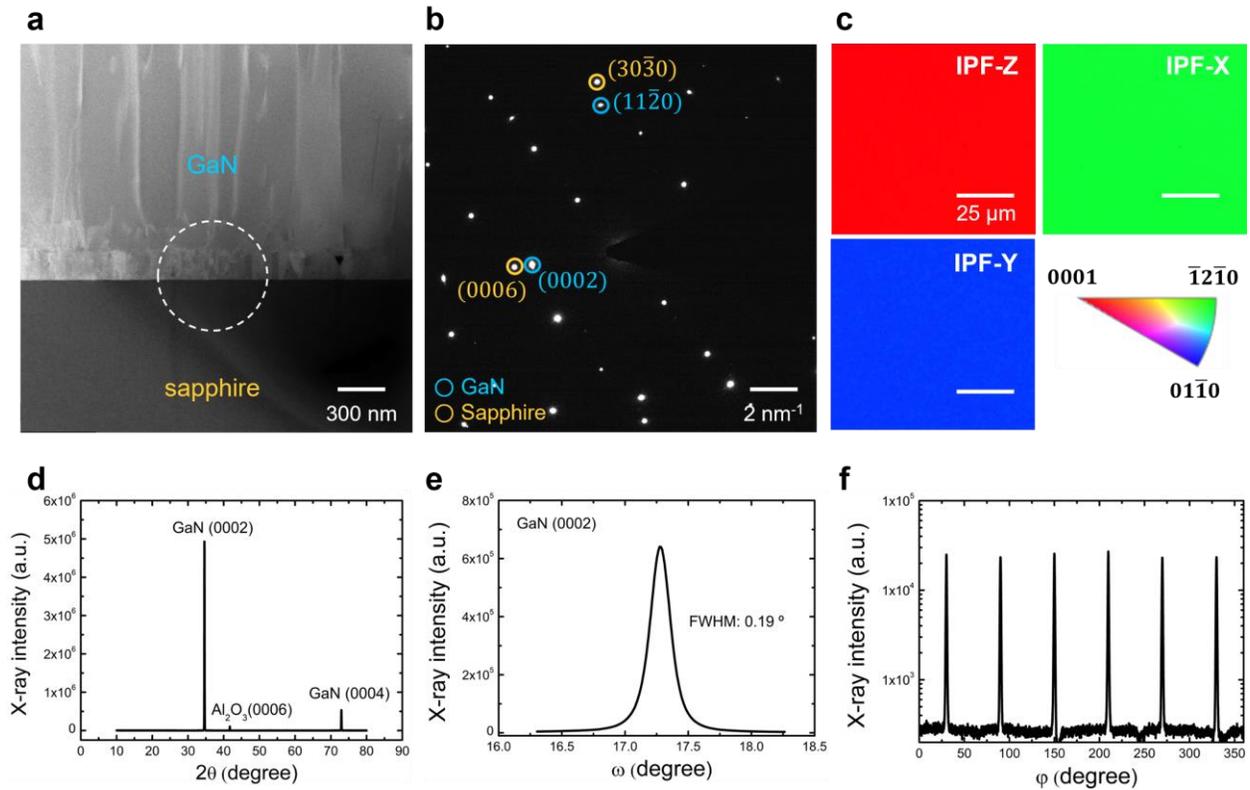

**Figure 4.** GaN thin films grown by remote epitaxy on a graphene-coated *c*-sapphire substrate by using a three-step growth method with the pulsed-flow NH$_3$ in the second growth step. (a) A low-magnification STEM image and (b) selected area electron diffraction (SAED) pattern at the interface between GaN and sapphire. (c) Electron backscatter diffraction (EBSD) inverse pole figure (IPF) maps, (d) X-ray diffraction (XRD) theta/2theta scan, (e) GaN(0002) XRD rocking curve, and (f) XRD azimuthal scan of the as-grown GaN thin films on graphene-coated *c*-sapphire.

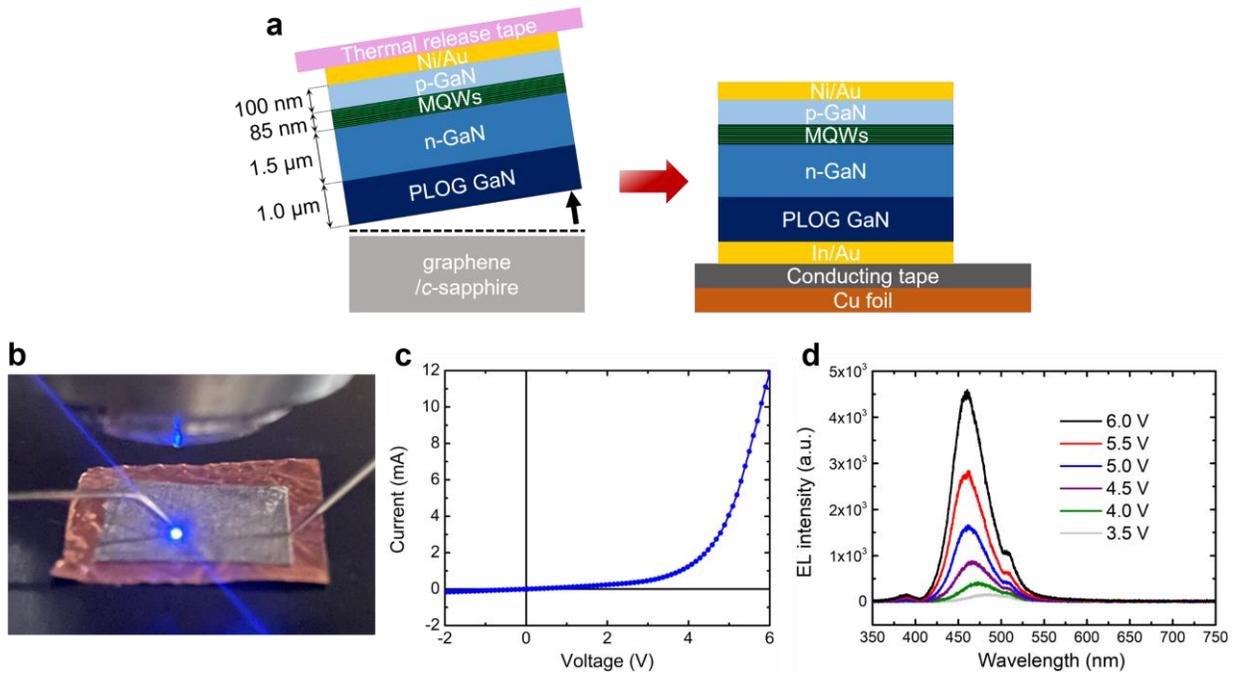

**Figure 5.** (a) Schematic of the GaN thin-film LEDs device structure fabricated on copper foil. (b) Photograph capturing light emission from GaN thin-film LEDs. (c) *I-V* characteristic curve of transferred LEDs with positive bias applied to Ni/Au and negative bias applied to the copper foil. (d) Room-temperature electroluminescence (EL) spectra at various applied bias voltage levels.

## ASSOCIATED CONTENT

**Supporting Information**

The supporting information is available free of charge.

Effect of $O_2$ plasma treatment on graphene and subsequent GaN buffer layer growth, annealing test on graphene/*c*-sapphire, metallic stressor layer-assisted exfoliation test, step-by-step exfoliation and Raman measurement, remote heteroepitaxy of GaN grown on a graphene-coated *c*-sapphire substrate, and additional cross-sectional STEM images of GaN grown by the $NH_3$ pulsed-flow method (PDF)

## AUTHOR INFORMATION

**Corresponding Author**

Gyu-Chul Yi − Department of Physics and Astronomy, Seoul National University, Seoul 08826, Republic of Korea; Email: gcyi@snu.ac.kr

**Authors**

Seokje Lee − Department of Physics and Astronomy, Seoul National University, Seoul 08826, Republic of Korea

Muhammad S. Abbas – Department of Physics, Sungkyunkwan University College of Natural Science, Suwon 16419, Republic of Korea

Dongha Yoo − Department of Physics and Astronomy, Seoul National University, Seoul 08826, Republic of Korea

Keundong Lee − Department of Electrical and Computer Engineering, University of California San Diego, La Jolla, CA 92093, United States

Tobiloba G. Fabunmi − Department of Physics and Astronomy, Seoul National University, Seoul 08826, Republic of Korea

Eunsu Lee − Department of Physics and Astronomy, Seoul National University, Seoul 08826, Republic of Korea

Imhwan Kim − Department of Physics and Astronomy, Seoul National University, Seoul 08826, Republic of Korea

Daniel Jang − SKKU Advanced Institute of Nano Technology, Sungkyunkwan University, Suwon 16419, Republic of Korea

Sangmin Lee − Department of Materials Science and Engineering, Seoul National University, Seoul 08826, Republic of Korea

Jusang Lee − Department of Materials Science and Engineering, Seoul National University, Seoul 08826, Republic of Korea

Ki-Tae Park − Department of Mechanical Engineering, Seoul National University, Seoul 08826, Republic of Korea

Changgu Lee – SKKU Advanced Institute of Nano Technology, Sungkyunkwan University, Suwon 16419, Republic of Korea; School of Mechanical Engineering, Sungkyunkwan University College of Engineering, Suwon 16419, Republic of Korea

Miyoung Kim – Department of Materials Science and Engineering, Seoul National University, Seoul 08826, Republic of Korea

Yun Seog Lee − Department of Mechanical Engineering, Seoul National University, Seoul 08826, Republic of Korea

Celesta S. Chang – Department of Physics and Astronomy, Seoul National University, Seoul 08826, Republic of Korea**Notes**

The authors declare no competing financial interest.

ACKNOWLEDGMENT

We acknowledge the support provided by the Science Research Center (SRC) for Novel Epitaxial Quantum Architectures and the National Research Foundation (NRF) of Korea (NRF-2021R1A5A1032996). This research was also supported by grants NRF-2020R1A2C2014687, NRF-2022R1A2C3007807, and NRF-2020R1C1C1005880 from the NRF of Korea. Additionally, we appreciate the Brain Korea 21-Plus Program, the Institute of Applied Physics (IAP), Research Institute of Advanced Materials (RIAM), Institute of Advanced Machines and Design (IAMD), and the Creative-Pioneering Researchers Program through the Institute of Engineering Research at Seoul National University.20

# Supporting information for

# Pulsed-mode metalorganic vapor-phase epitaxy of GaN on graphene-coated *c*-sapphire for freestanding GaN thin films

*Seokje Lee[1], Muhammad S. Abbas[2], Dongha Yoo[1], Keundong Lee[3], Tobiloba G. Fabunmi[1], Eunsu Lee[1], Imhwan Kim[1], Daniel Jang[4], Sangmin Lee[5], Jusang Lee[5], Ki-Tae Park[6], Changgu Lee[4,7], Miyoung Kim[5], Yun Seog Lee[6], Celesta S. Chang[1], Gyu-Chul Yi[1]\**

AUTHOR ADDRESS

[1]Department of Physics and Astronomy, Seoul National University, Seoul 08826, Republic of Korea

[2]Department of Physics, Sungkyunkwan University College of Natural Science, Suwon 16419, Republic of Korea

[3]Department of Electrical and Computer Engineering, University of California San Diego, La Jolla, CA 92093, United States

[4]SKKU Advanced Institute of Nano Technology, Sungkyunkwan University, Suwon 16419, Republic of Korea

[5]Department of Materials Science and Engineering, Seoul National University, Seoul 08826, Republic of Korea

[6]Department of Mechanical Engineering, Seoul National University, Seoul 08826, Republic of Korea

[7]School of Mechanical Engineering, Sungkyunkwan University College of Engineering, Suwon 16419, Republic of Korea

*Corresponding author: gcyi@snu.ac.kr

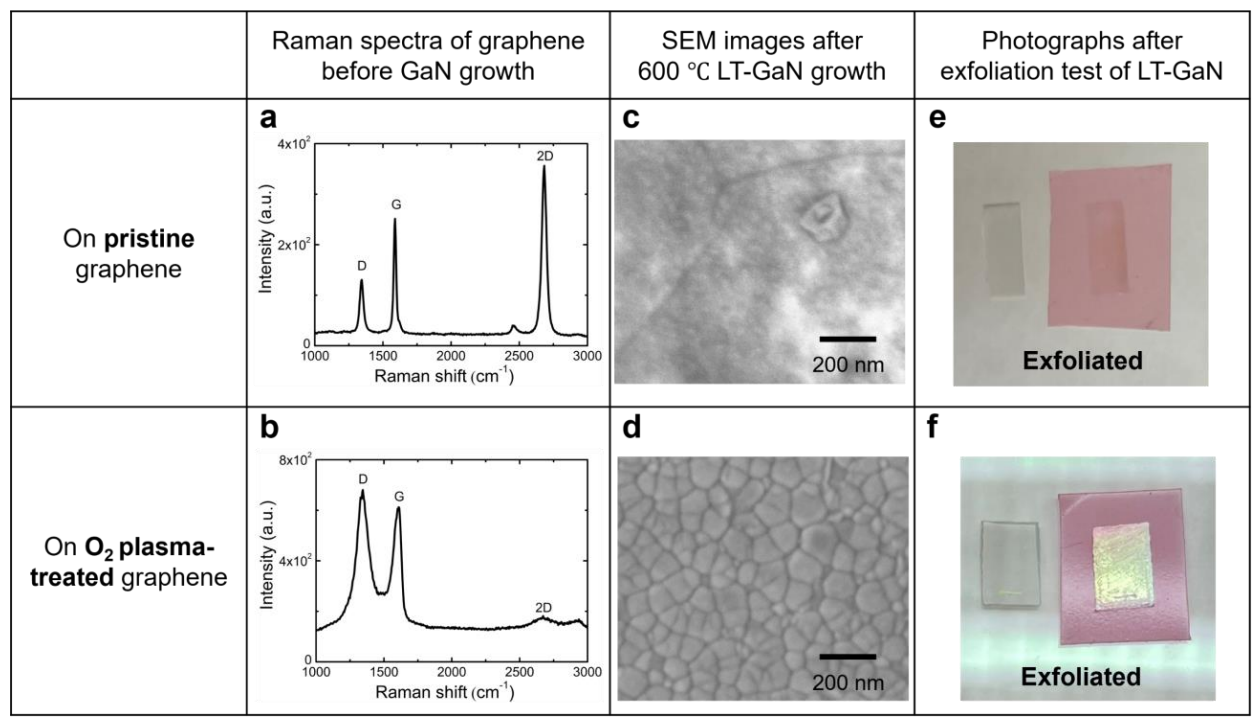

**Figure S1.** Effect of O₂ plasma treatment on graphene and subsequent GaN buffer layer growth. Raman spectra of (a) pristine graphene on *c*-sapphire and (b) O₂ plasma-treated graphene on *c*-sapphire, and SEM images of low-temperature GaN (LT-GaN) buffer layer grown on (c) pristine graphene/*c*-sapphire and (d) O₂ plasma-treated graphene/*c*-sapphire. Photographs depict the exfoliation tests for LT-GaN grown on (e) pristine graphene/*c*-sapphire and (f) O₂ plasma-treated graphene/*c*-sapphire.

We created nucleation sites for GaN on graphene by inducing dangling bonds on the graphene surface using a mild O₂ plasma treatment (3 seconds under 40W plasma power) before the GaN growth process. Prior to the O₂ plasma treatment, Raman spectra of pristine graphene clearly displayed D, G, and 2D peaks (Figure S1a). However, after the treatment, Raman spectroscopy confirmed the formation of dangling bonds, as evidenced by the significant increase in the D peak and the broadening and decrease of the 2D peak (Figure S1b). We then grew a low-temperature GaN (LT-GaN) layer on both the pristine and O₂ plasma-treated graphene using MOVPE at 600 °C for 10 min. Scanning electron microscope (SEM) image shows that only a low-density of GaN buffer islands were observed on the pristine graphene (Figure S1c), while GaN buffer islands with diameters of 30-130 nm were densely grown on the $O_2$ plasma-treated graphene surface (Figure S1d). In addition, we exfoliated two LT-GaN samples without metal stressor deposition using thermal release tape and found that both samples were completely exfoliated, regardless of whether they were on pristine or $O_2$ plasma-treated graphene substrates (Figure S1e,f). These results demonstrate that mild $O_2$ plasma treatment can be used to enhance the nucleation of GaN growth on graphene without causing severe graphene etching that could lead to a failure in GaN exfoliation.

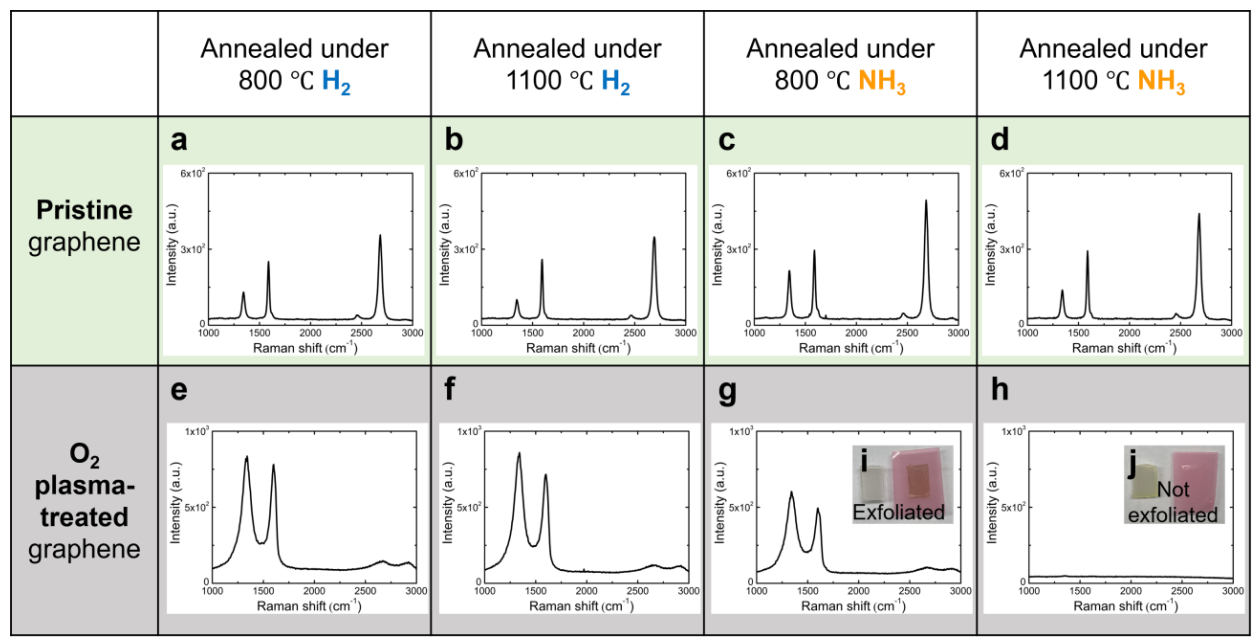

**Figure S2.** Annealing test on graphene/*c*-sapphire. Raman spectra of pristine graphene on *c*-sapphire substrates annealed for 10 min under 100 torr: (a) $H_2$ at 800 °C, (b) $H_2$ at 1100 °C, (c) $NH_3$ at 800 °C, and (d) $NH_3$ at 1100 °C. Raman spectra of $O_2$ plasma-treated graphene on *c*-sapphire substrates annealed for 10 min under 100 torr: (e) $H_2$ at 800 °C, (f) $H_2$ at 1100 °C, (g) $NH_3$ at 800 °C, and (h) $NH_3$ at 1100 °C. Photographs after exfoliation tests for LT-GaN grown on $O_2$ plasma-treated graphene annealed for 10 min under 100 torr: (i) $NH_3$ at 800 °C and (j) $NH_3$ at 1100 °C.

We conducted a systematic annealing test on both pristine graphene and $O_2$ plasma-treated graphene on *c*-sapphire substrates to investigate the factors contributing to graphene damage. The annealing was carried out for 10 min at a reactor pressure of 100 torr, with varying temperatures (800 °C and 1100 °C) and ambient gas flows ($H_2$ 2000sccm and $NH_3$ 2000sccm). Following the annealing test, we examined the graphene conditions using Raman spectroscopy. For pristine graphene, no significant changes were observed in the Raman intensity under all annealing conditions, showing clear D, G, and 2D peaks of graphene (Figure S2a-d). For the $O_2$ plasma-treated graphene, $H_2$-flow annealing did not affect the Raman spectra at either 800°C (Figure S2e) or 1100°C (Figure S2f). However, the D and G intensity of $O_2$ plasma-treated graphene slightly decreased after annealing under $NH_3$ flow at 800 °C (Figure S2g). Moreover, the Raman signal of $O_2$ plasma-treated graphene disappeared after annealing under $NH_3$ flow at 1100°C (Figure S2h). To further confirm the presence of graphene, we grew LT-GaN on $O_2$ plasma-treated graphene samples annealed under $NH_3$ flow at 800 °C and 1100 °C. When attempting to exfoliate these LT-GaN samples using thermal release tape, LT-GaN grown on $O_2$ plasma-treated graphene annealed under $NH_3$ flow at 800 °C was completely exfoliated (Figure S2i), while the sample annealed under $NH_3$ flow at 1100 °C was not exfoliated except at the edge (Figure S2j). These results suggest that $NH_3$ damage the $O_2$ plasma-treated graphene, especially at high-temperature conditions, likely due to reactive sites generated by the $O_2$ plasma treatment on graphene.

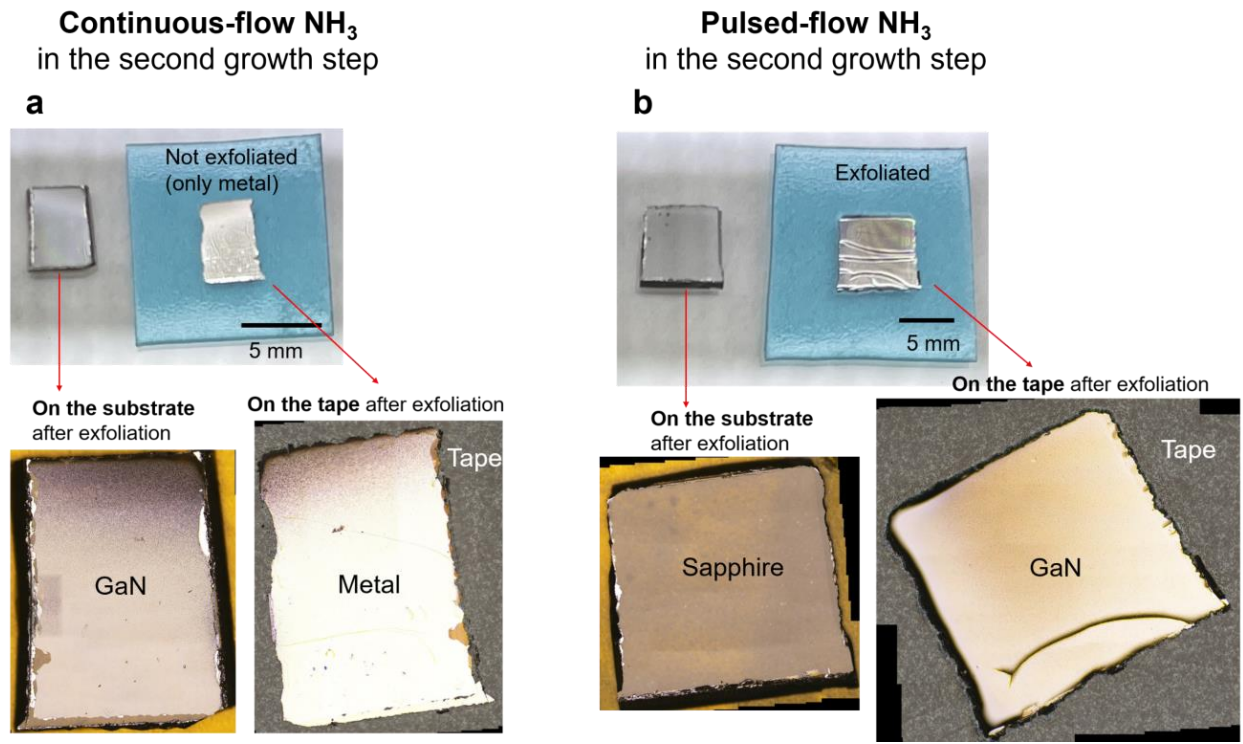

**Figure S3.** Metallic stressor layer-assisted exfoliation test. Photographs and optical microscope images of exfoliation test using Ti/Ni metal stressor deposition for GaN thin films on graphene-coated sapphire grown by using (a) a continuous-flow NH$_3$ and (b) a 2s-on/2s-off pulsed-flow NH$_3$ in the second growth step.

We initially used only thermal release tape to exfoliate the GaN thin films grown on the graphene-coated sapphire substrates. We observed that the GaN film grown by pulsed-flow NH$_3$ in the second growth step was completely exfoliated, while the film grown by continuous-flow NH$_3$ in the second growth step was not exfoliated at all. To investigate whether the film required higher stress for exfoliation, we deposited a 50nm Ti and 1μm Ni metallic stressor layer on the surface of the GaN films using sputtering before applying the thermal release tape. However, the GaN film grown by continuous-flow NH$_3$ was still not exfoliated, and only the metallic stressor layers were peeled off from the GaN surface (Figure S3a). In contrast, the GaN film grown by pulsed-flow $NH_3$ was exfoliated over 99% of its area (Figure S3b).

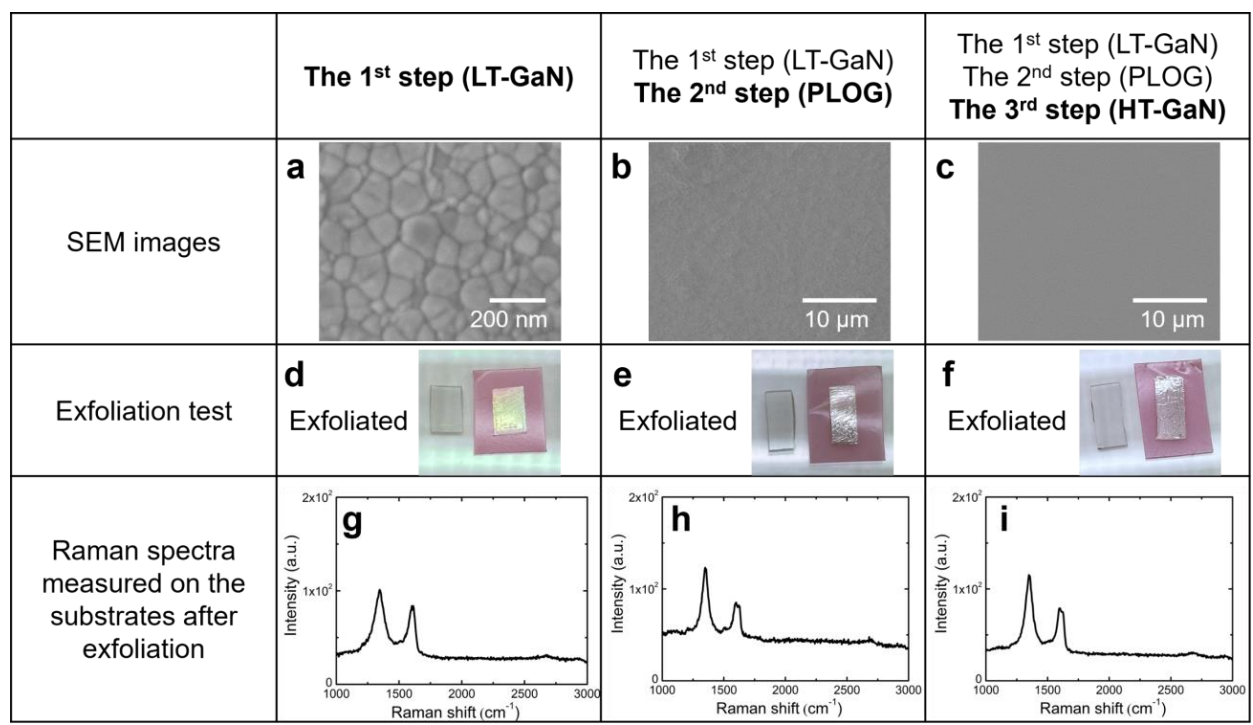

**Figure S4.** Step-by-step exfoliation and Raman characterization. SEM images of GaN grown on graphene-coated *c*-sapphire substrates, including (a) LT-GaN after the first growth step, (b) pulsed epitaxial lateral overgrowth (PLOG) GaN after the second growth step, and (c) high-temperature GaN (HT-GaN) after the third growth step. Exfoliation test results of (d) LT-GaN, (e) PLOG GaN, and (f) HT-GaN from the graphene-coated sapphire substrates. Raman spectra measured on the substrates after (g) LT-GaN, (h) PLOG GaN, and (i) HT-GaN exfoliation.

The presence of graphene was confirmed through exfoliation test and Raman spectroscopy after each growth step. Scanning electron microscopy (SEM) images in Figure S4a-c depict the surface morphology of the GaN samples grown on graphene-coated *c*-sapphire substrates after each growth step of the first LT-GaN growth, the second pulsed epitaxial lateral overgrowth (PLOG), and the third high-temperature GaN (HT-GaN) growth. Successful exfoliation of every GaN sample was achieved (Figure S4d-f). Raman spectroscopy was also carried out on every original graphene-coated sapphire substrate surface after exfoliation, which confirmed the presence of D and G Raman peaks with a weak 2D peak of graphene for all the original substrates (Figure S4g-i). Moreover, no significant change in the Raman peaks of graphene was observed after each growth step. These results suggest that the PLOG GaN layer grown under a pulsed flow of $NH_3$ protected the underlying graphene during both the PLOG and HT-GaN steps, leading to the successful preservation of graphene throughout the GaN growth process.

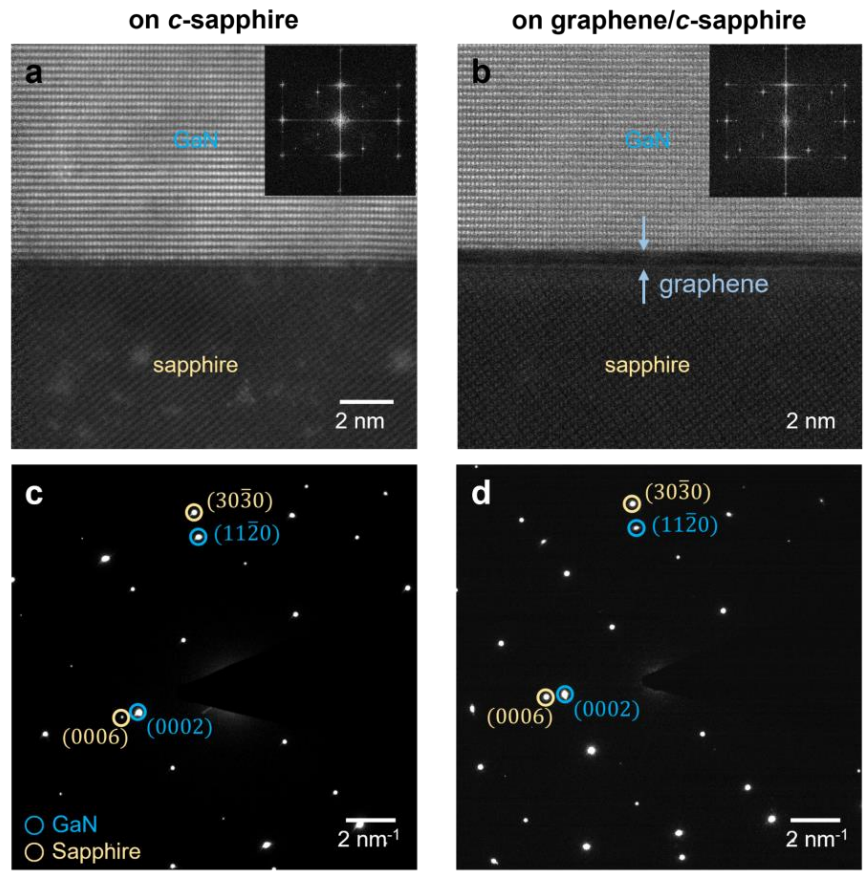

**Figure S5.** Remote heteroepitaxy of GaN grown on a graphene-coated *c*-sapphire substrate. Cross-sectional scanning transmission electron microscopy (STEM) high-angle annular dark-field (HAADF) images with fast Fourier transform (FFT) patterns (insets) of GaN films grown on (a) bare *c*-sapphire and (b) graphene-coated *c*-sapphire. Selected area electron diffraction (SAED) patterns of the GaN/sapphire interface of GaN films grown on (c) bare *c*-sapphire and (d) graphene-coated *c*-sapphire.

We investigated the hetero remote epitaxy of GaN on a graphene-coated *c*-sapphire substrate by comparing samples simultaneously grown on bare and graphene-coated *c*-sapphire substrates in the same growth batch. The cross-sectional scanning transmission electron microscopy (STEM) high-angle annular dark-field (HAADF) images and corresponding fast Fourier transform (FFT) patterns of both samples reveal the same atomic arrangement of GaN aligned with the *c*-axis and epitaxial relationship between GaN and *c*-sapphire (Figure S5a, b). The selected area electron diffraction (SAED) revealed that GaN grown on bare *c*-sapphire showed GaN (0002) // sapphire(0006) and GaN(11$\bar{2}$0) // sapphire(30$\bar{3}$0) heteroepitaxial relationship (Figure S5c). The same heteroepitaxial relationship was observed in the SAED pattern of GaN grown on graphene-coated *c*-sapphire (Figure S5d), indicating that GaN thin films were grown while maintaining the heteroepitaxial relationship with the *c*-sapphire substrate despite the presence of graphene.

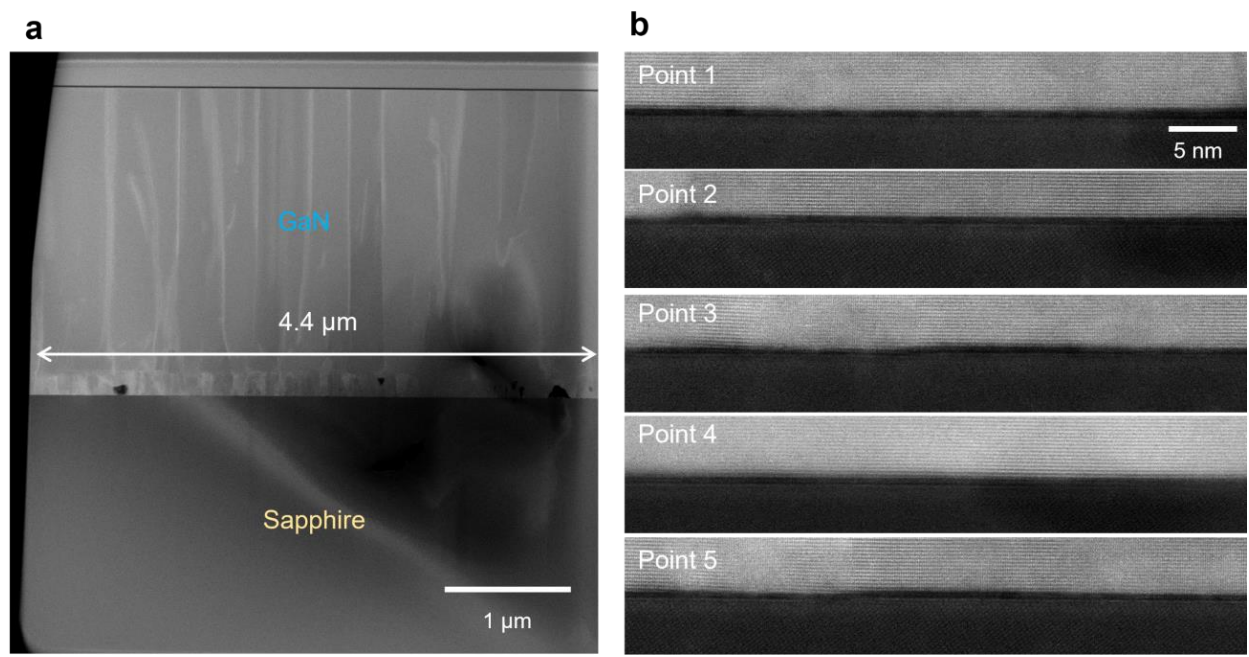

**Figure S6.** Additional cross-sectional STEM images of GaN grown by the $NH_3$ pulsed-flow method. STEM-HAADF images of GaN thin films grown on a graphene-coated *c*-sapphire substrate using the three-step growth method with a PLOG step. (a) Low-magnification STEM-HAADF image of the entire cross-sectional region with a length of 4.4 μm. (b) High-magnification STEM-HAADF images of five measured points from the entire region.

We used cross-sectional STEM to investigate the GaN/graphene/sapphire interfaces of GaN thin films that were grown on a graphene-coated sapphire substrate using the three-step growth method with PLOG method. By scanning the entire 4.4 μm cross-section region (Figure S6a), we did not observe any holes through which GaN directly contacted the sapphire substrate. Figure S6b displays the STEM-HAADF images of five representative measured points from the entire region, and all show a distinct gap between GaN and the sapphire substrate, indicating the presence of graphene.